\documentclass[a4paper,twoside]{article}
\usepackage{graphicx}
\usepackage{amsmath}
\usepackage{amssymb}
\newcommand{\affil}[1]{$^{\rm #1}$}
\usepackage[authoryear]{natbib}
\bibpunct{(}{)}{;}{a}{}{,}

\date{} 
\title{\large\bf\flushleft Pre-mainsequence stellar evolution in $\pmb{N}$-body models}
\author{\parbox{\textwidth}{\flushleft
\vspace{-0.5cm}
{\it Anna D. Railton\affil{A}, Christopher A. Tout\affil{A,B}, and Sverre J. Aarseth\affil{A}}\\
\vspace{0.4cm}
{\small \affil{A}\,Institute of Astronomy, The Observatories, Madingley Road, Cambridge
CB3 0HA, England}\\
{\small \affil{B}\,Email: cat@ast.cam.ac.uk}}}
\begin{document}
\twocolumn[
\begin{changemargin}{.8cm}{.5cm}
\begin{minipage}{.9\textwidth}
\vspace{-1cm}
\maketitle
%
%
\small{\bf Abstract:}
We provide a set of analytic fits to the radii of pre-mainsequence
stars in the mass range $0.1<M/M_{\odot}<8.0$.  We incorporate
the formulae in $N$-body cluster models for evolution from the
beginning of pre-main sequence.  In models with 1\,000 stars and high
initial cluster densities, pre-mainsequence evolution causes roughly
twice the number of collisions between stars than in similar models with
evolution begun only from the zero-age main sequence.  The collisions
are often all part of a runaway sequence that creates one relatively massive
star.


\medskip
\medskip
\end{minipage}
\end{changemargin}
]
\small

\section{Introduction}

Existing $N$-body simulations of stellar clusters usually begin with
all stars on the zero-age main sequence (ZAMS).  Our motivation to
include pre-mainsequence (preMS) evolution in such models goes beyond the desire for
completeness. Given the upper mass limit ($M>10\,M_\odot$) attributed
to conventional star formation \citep{2000prpl.conf..327S} and because
massive stars are mainly found in the denser central regions of star
clusters \citep[e.g.][]{1998ApJ...492..540H},
\citet{1998MNRAS.298...93B} first suggested collisions between preMS
stars as a means to create massive stars.  PreMS
stars are larger than their main-sequence counterparts and, although
the phase is short-lived, there is a possibility that the increased
collision likelihood has an effect.

The journey from molecular gas cloud to protostar to the \rm{ZAMS} is
complicated.  If random motions owing to MHD turbulence are sustained
inside a molecular cloud, density inhomogeneities form and quickly
collapse under their own gravity to become protostars
\citep*{2005fost}. Typically, when modelling preMS evolution, this
self-gravitating fragment of a protostellar cloud is where the models
begin \citep*{tout1999} and is where we define the zero age of the
pre-main sequence.

In the 1950s, Henyey constructed the first detailed numerical models
of young stars, assuming that preMS stars are radiatively stable
\citep*{1955PASP...67..154H}. This led to nearly horizontal tracks in
the Hertzsprung--Russell (H--R) diagram because the stars contract at almost constant
luminosity.  \citet{1961PASJ...13..450H} realised that the $\rm{H}^-$
opacity (and Kramer's opacity, when the temperature increases) forces
the young stars to radiate at nearly constant effective temperature
$T_{\rm{eff}}$ and so to follow nearly vertical tracks in an
H--R diagram.  These young stars are in a quasi-hydrostatic
equilibrium.  They contract on a time-scale much greater than the
free-fall time-scale \citep*{1961PASJ...13..450H}.  Although protostars
are initially homogeneous and isothermal, they collapse
non-homogeneously, creating high central density and temperature. It
is this that eventually gives rise to the suitable conditions for
hydrogen fusion in stars with $M>0.08\,M_{\odot}$. Below this mass,
the central conditions are never sufficient for hydrogen fusion and
the stars collapse to become brown dwarfs
\citep{1963ApJ...137.1121K,1963PThPh..30..460H}.

\begin{figure}
  \begin{center} \includegraphics[angle=-90,width=84mm]{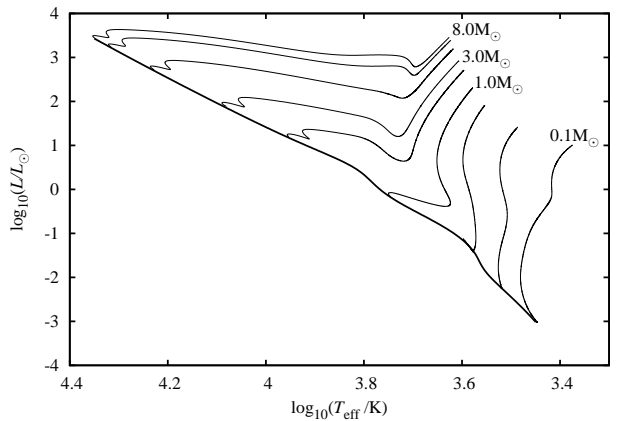}
  \caption{An H--R diagram showing the ZAMS in bold and the preMS
  tracks for a range of masses. The distinction between the vertical
  Hayashi track and horizontal Henyey track is apparent here. Data is
  taken from \citet*{Tout96} and \citet*{tout1999}.}  \label{fig:hrdiag}
  \end{center}
\end{figure}

Stars with $M \le 0.7\,M_{\odot}$ behave as though they are fully
convective and reach the ZAMS at the end of their Hayashi track. When $M
\ge 0.7\,M_{\odot}$, the star develops a radiative core which is large
enough that it evolves more like Henyey's initial models,
turning on to the near-horizontal Henyey track where contraction
continues at approximately constant luminosity. This behaviour can be
seen in Figure~\ref{fig:hrdiag}, which shows the calculated preMS tracks
modelled by \citet{tout1999}.  However \citet{2000prpl.conf..327S}
argue that stars more massive than $10\,M_{\odot}$ have their
contraction disrupted by both strong winds and radiation pressure and
thus the conventional theory of protostellar infall fails in this
case.  \citet{2007ARA&A..45..481Z} discussed three
competing mechanisms for massive star formation.  These are the turbulent core
model of \citet{2003ApJ...585..850M}, the competitive accretion
process of \citet{1997MNRAS.285..201B} and stellar collisions
and merging of stars during their preMS evolution or later.  It is this
third case that we investigate here.

An alternative model was investigated by \citet{baumgardt2011} and it
is interesting to compare our results with theirs.  They base their
pre-mainsequence evolution on models of accreting stars constructed by
\citet{bernasconi1996}.  The physics used by \citet{bernasconi1996} is
not very different from that used by \citet{tout1999} and indeed the
evolutionary tracks in the H--R diagram are very
similar once accretion has ceased and the stars contract down Hayashi
tracks.  The major difference between the models we present here and
those of \citet{baumgardt2011} is that all their stars begin as
protostellar cores of $0.1\,\rm M_\odot$ and accrete at a constant rate
until they arrive at a suitable mass to populate an IMF up to $15\,\rm
M_\odot$.  Their stars then undergo a preMS phase in which the
logarithm of their radii shrinks linearly with time until they reach
the ZAMS, whereupon they begin main-sequence evolution.  Instead we
start with a set of pre-mainsequence stars with masses populating a similar
IMF up to $8\,\rm M_\odot$ that begin their evolution already fully
grown on Hayashi tracks.  Our fitting formulae are a little more
complicated because we attempt to to fit both the Hayashi and Henyey
phases of pre-mainsequence evolution together.  Because of this our
stars remain larger for longer (see section~\ref{conc}) after any
accretion has ceased.  We include this behaviour in cluster models
with a variety of initial conditions to test its effect.

\section{Models}

We use the {\sc nbody6}\footnote{This code is available to download at
http://www.ast.cam.ac.uk/research/nbody.} \citep{aarseth1999}.  This code incorporates
stellar evolution and binary interaction by empirical formulae fitted
to detailed stellar models.  Two major additions to these standard
packages are required.  First we must have a reasonable empirical
representation of how preMS stars, and particularly their radii, evolve.
Secondly we must include a model of what happens when they collide.

\subsection{Parameterization of the preMS evolution}

\label{sec:prems}

We construct a set of fitting functions for the preMS evolution of
stars with masses $M$ in the range $0.1<M/M_{\odot}<8.0$.  We choose
to find analytical fits rather than to tabulate data in our models
because this approach is in line with the parameterized treatment of
stellar evolution in the other parts of the $N$-body code.  Such fits
also have the advantage that they are continuous and differentiable
and this makes modelling preMS evolution with accretion simpler
because the fits are smooth functions of mass.  We began with detailed
stellar models originally constructed by \citet{tout1999} with
the Cambridge {\sc stars} code \citep{eggleton} described by
\citet*{STARS}.  We used data for nine stars with solar metallicity
$Z=0.02$ and masses $M\in\lbrace0.1, 0.2 ,0.5 ,1.0 ,2.0 ,3.0, 5.0,
7.0, 8.0\rbrace\,M_{\odot}$.  These masses were chosen because stars
with $M<0.08\,M_{\odot}$ never meet the ZAMS and for $M>8\,M_{\odot}$
the time-scale of preMS evolution is short enough that it is safe to
neglect it.  The time taken for our $8\,M_\odot$ star to contract from
about $90\,\rm R_\odot$ to the main sequence is about $3\times
10^5\,$yr.  It takes only the first $1.25\times 10^4\,$yr to contract
to $40\,\rm R_\odot$, a radius it does not exceed again until it
becomes a red giant.  It then spends about $5\times 10^4\,$yr around
$36\,\rm R_\odot$ while it burns its supply of deuterium fuel and then
loses its convective envelope, so moving from the Hayashi to the
Henyey phase.  In section~\ref{seccollfreq} we compare this with the
expected collision time-scale in the densest
clusters that we model.  We find that neglect of preMS evolution for
stars above $8\,\rm M_\odot$ is reasonable for these clusters but
would not be for much denser systems.

\begin{figure}
\begin{center}
\includegraphics[width=84mm]{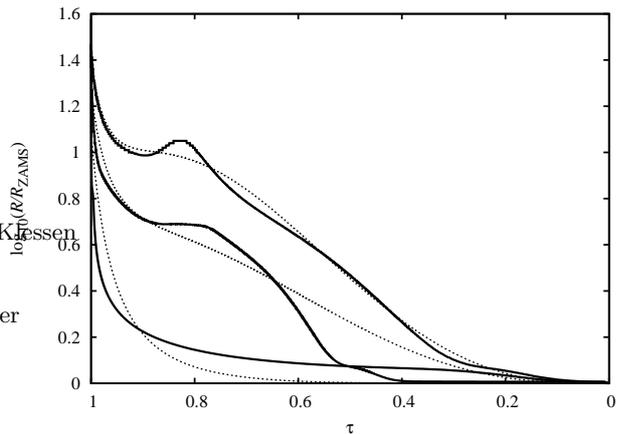}
\caption{
Fitted preMS radii for $1.0\,M_{\odot},$ $5.0\,M_{\odot}$ and
$8.0\,M_{\odot}$ (bottom to top) at $\tau = 0.8$.
The solid lines are the detailed models and the dashed lines
the fits.  Note
the reversed scale in $\tau$, with $\tau=1$ at the beginning of the
preMS track and $\tau=0$ when the star reaches the ZAMS.}
\label{fig:fits}
\end{center}
\end{figure}

\begin{table*}
\begin{center}
\caption{Coefficients for equation (\ref{eq:r})}
\smallskip
\begin{tabular}{l l l l l l l l l l}
\hline
$i$&$\alpha^{(1)}_{i}$&$\alpha^{(2)}_{i}$&$\alpha^{(3)}_{i}$&$\beta^{(1)}_{i}$&$\beta^{(2)}_{i}$&$\beta^{(3)}_{i}$&$\gamma^{(1)}_{i}$&$\gamma^{(2)}_{i}$&$\gamma^{(3)}_{i}$ \\
\hline
0&0&          -4.00772&\phantom{-}1.60324&0&\phantom{-}8.5656&          -4.56878&\phantom{-}0.07432&          -4.50678&\phantom{-}3.01153\\
1&0&\phantom{-}4.00772&\phantom{-}2.20401&0&	      -8.5656&          -4.05305&          -0.09430&\phantom{-}4.56118&\phantom{-}1.85745\\
2&0&\phantom{-}0      &          -0.60433&0&\phantom{-}0     &\phantom{-}1.24575&\phantom{-}0.07439&\phantom{-}0      &          -0.64290\\
3&0&\phantom{-}0      &\phantom{-}0.05172&0&\phantom{-}0     &          -0.10922&\phantom{-}0      &\phantom{-}0      &\phantom{-}0.05759\\
\hline
\end{tabular}
\label{tab:rcoef}
\end{center}
\end{table*}

Given the general difficulty of deciding the birth time of a star and
hence its age, we measure the age of preMS stars backwards from the ZAMS. 
We therefore devise a slightly artificial preMS
time-scale $\tau_{\rm{preMS}}$ by taking the time from which the
\citet{tout1999} models begin until their radius meets the ZAMS
value.  To do this we use the analytic \rm{ZAMS} radius formula of
\citet*{Tout96}.  It is a function of mass and metallicity but all our
fits are made for solar metallicity $Z=0.02$, for which differences
from the detailed ZAMS models
reported by \citet{STARS} are at most 1.2\%. It is imperative that these particular
errors are small because all other formulae calculate properties
relative to these ZAMS values. Using a least squares method, we fit
these times as a function of mass to get the preMS time-scale in the form
\begin{multline}
\log_{10}\left(\frac{\tau_{\rm{preMS}}}{\rm{yr}}\right)=43.6-35.8\left(\frac{M}{M_{\odot}}\right)^{0.015}\\
\times\exp{\left[3.96 \times 10^{-3}\left(\frac{M}{M_{\odot}}\right)\right]}
\label{eq:tzams}
\end{multline}
for our mass range.  Next we define a scaled time 
\begin{equation}
  \tau=1-t/\tau_{\rm{preMS}}
  \label{eq:tau}
\end{equation}
such that time runs from $\tau=1$ at the beginning of the preMS track
to the ZAMS at $\tau=0$ and hence $\tau$ is strictly in the range
$\tau \in [0,1]$. These functions take the form
\begin{equation}
  R=R_{\rm{ZAMS}}\left(M\right)10^{f(\tau)},
    \label{eq:r}
\end{equation}
where
\begin{equation}
f(\tau)=\frac{\alpha\tau^3+\beta\tau^4+\gamma\tau^5}{1.05-\tau}.
\end{equation}
\noindent For each of the nine models we obtain a best fit for
$\alpha$, $\beta$ and~$\gamma$ and then fit these three
coefficients as functions of mass. Several iterations were necessary
to converge on good formulae. The coefficients $\alpha$, $\beta$
and $\gamma$ are all well represented as piecewise cubics in mass of the forms
\begin{align}
  \alpha^{(j)}(\tfrac{M}{M_{\odot}})&=\sum_{i=0}^3\alpha_i^{(j)}(\tfrac{M}{M_{\odot}})^i, \notag\\
  \beta^{(j)}(\tfrac{M}{M_{\odot}})&=\sum_{i=0}^3\beta_i^{(j)}(\tfrac{M}{M_{\odot}})^i\\
  \text{and}\quad\gamma^{(j)}(\tfrac{M}{M_{\odot}})&=\sum_{i=0}^3\gamma_i^{(j)}(\tfrac{M}{M_{\odot}})^i, \notag
\end{align}
where
\begin{equation}
  j=
  \begin{cases}
    1 & M \leq 1\,M_{\odot},\\
    2 & 1\,M_{\odot} < M < 2\,M_{\odot}, \\
    3 & M \geq 2\,M_{\odot}
   \end {cases}
\end{equation}
and $\alpha^{(j)}_i$, $\beta^{(j)}_i$ and $\gamma^{(j)}_i$ are listed
in Table \ref{tab:rcoef}.  Our fits (illustrated in
Figure~\ref{fig:fits}) are all physically reasonable.  They
incorporate both the Hayashi and Henyey tracks.  Sadly the fits are
not as accurate as we may have liked.  However we do not wish to make
the formulae excessively complicated for what is a rather short phase
of evolution.  Our biggest deviations are around $5\,\rm M_\odot$,
shown in the figure, for which we overestimate the radius by a factor of
between about $1.25$ and~$1.5$ for around 20\% of its pre-mainsequence
lifetime.  Averaged over all masses we tend to underestimate as much
as overestimate.

\subsection{Treatment of collisions}

In addition to the treatment of stellar evolution we must also
consider collisions of preMS stars with themselves and stars of other
types. We model a collision between two preMS stars as if both are
$n=3/2$ polytropes because stars on Hayashi tracks are fully
convective and most collisions occur whilst stars are largest.  The
gravitational energy of an $n=3/2$ polytrope of mass $M$ and
radius $R$ is
\begin{equation}
\Omega=-\frac{6}{7}\frac{GM^2}{R}.
\end{equation}
The internal
energy of a star in virial equilibrium is
\begin{equation}
U=-\frac{1}{3(\gamma-1)}\Omega.
\end{equation}
This is a reasonable approximation
to make because the preMS stars are in quasi-hydrostatic
equilibrium.  For a perfect gas, $\gamma=5/3$,
so
\begin{equation}
U=\frac{3}{7}\frac{GM^2}{R}.
\end{equation}

We consider a collision between two preMS stars with masses $M_1$
and $M_2$, radii $R_1$ and $R_2$ and total mass $M=M_1+M_2$.  We assume
that colliding stars merge while a small fraction of the total mass $\xi M$ is lost from the
cluster \citep{1993A&A...272..430D}, and let energy be conserved during
the collision. The two stars are moving with high velocity when they
collide but the kinetic energy at infinity, or at the apogee of an
eccentric binary orbit, in their relative orbits can be
neglected. The initial velocity of the colliding stars would need to
be of order $10^3\,\rm{km}\,\rm{s}^{-1}$ to have kinetic energy
comparable to the energy lost with $\xi M$, whereas in globular
clusters and star-forming regions the velocities are typically
$10\,\rm{km}\,\rm{s}^{-1}$
\citep[e.g.][]{2010ARA&A..48..431P}. Conserving the internal energy
and ignoring this orbital contribution, we find that the radius
$R_{\ast}$ of the new coalesced star is

\begin{equation}
R_{\ast}=\frac{7M^2}{3}\left(\frac{M_1^2}{R_1}+\frac{M_2^2}{R_2}\right)^{-1}(3-4\xi)(1-\xi)
\end{equation}
and its mass is

\begin{equation}
M_{\ast} = M_1 + M_2 - \xi M.
\end{equation}

It is difficult to estimate mass loss in any stellar collision but,
for MS stars, SPH calculations by \citet*{1987ApJ...323..614B} indicate
about~10\%.  Thus $\xi=0.1$. \citet{2005ApJ...627..277L} also found
that a mass loss of a few~per cent of the total mass per collision with
a preMS star is a reasonable approximation.  We invert the
$R=R(M_{\ast},\tau)$ function~(\ref{eq:r}) to find the rejuvenated age at which to
restart evolution so that the star continues to contract from its new radius.

The outcome of a preMS/MS collision depends on the mass of the MS
star. Below $0.7\,M_{\odot}$ the MS star is still mostly convective
and, to a good approximation, the collision remnant can be modelled as
above. At higher masses, the MS star has very little convective
envelope and a dense convective core appears above $1.1\,M_{\odot}$.
Mass is therefore most likely to be accreted on to the surface of the
star but not mixed as above and evolution would not restart on the
preMS. All other collisions with different stellar types are treated
simply as accretion, so the non-preMS stellar type is kept upon
collision and the mass of the star is simply increased.  Such prescription
is again supported by \citet{2005ApJ...627..277L} who showed
that the collision product, when the collision involves a preMS star,
does not depend strongly on the impact parameter nor the initial
velocity.

\section{Results}
\label{sec:results}

Our initial conditions are all \citet{1911MNRAS..71..460P} models in virial equilibrium
with no primordial binary stars. There is no
interstellar gas nor mass segregation.  Our stellar evolution is as
described in section~\ref{sec:prems} and by \citet{hurley2000} and \citet{hurley2002}.  We use the
code {\sc nbody6} \citep{aarseth1999}.  To explore a didactic
set of $N$-body models, we keep the initial
number of stars $N$, the initial mass function for the stars and the
time-scale of evolution constant in our series of models.  We vary only
the half-mass radius $R_{0.5}$, the radius within which half the mass
of the cluster is contained.  We fix $N$ to be $1\,000$.  For the
masses of the stars we use
a modified Kroupa initial mass function
\citep{Kroupa01}, with $M_{\rm{min}}=0.1\,M_{\odot}$,
$M_{\rm{max}}=4.0\,M_{\odot}$ and $\overline{M}=0.4\,M_{\odot}$.
We chose the length-scale $\overline{R}\in\{0.02,0.05,0.1,0.2,0.4\}\rm{pc}$.
Our half-mass densities $\overline{\rho}_{0.5}$ range from $10^3$ to~$10^7\,M_{\odot}\,\rm{pc}^{-3}$.
The half-mass radius $R_{0.5}\approx 0.8\overline{R}$.

Because our preMS fitting is restricted to the mass range
$0.1<M/M_{\odot}<8.0$ (equation~\ref{eq:r}), if (a) different IMF
parameters were chosen with $M_{\rm{max}}\geq 8.0\,M_{\odot}$ or (b)
several stars collide to form a coalesced star with $M_{\ast}\geq
8.0\,M_{\odot}$ then these high-mass stars evolve from the ZAMS. This
approximation is justified because the preMS evolution of stars of
these high masses is so rapid that neglecting it loses little
information (see section~\ref{seccollfreq}).  We evolved each cluster
for $t=10\,\rm{Myr}$, approximately $\tau_{\rm{preMS}}$ for a
$0.5\,M_{\odot}$ star, with the logic that most interesting preMS
behaviour would have occurred by this time.  We made ten models for
each of the length-scales, both with, all the stars started at the top
of their Hayashi tracks at $\tau=1$, and without, all stars started on
the ZAMS at $\tau=0$, preMS evolution.  We shall hereinafter call
these \emph{preMS} and \emph{ZAMS} runs respectively.

\subsection{Densities}

\begin{table*}
\begin{center}
\caption{Evolution of the half-mass density
$\overline{\rho}_{0.5}=M_{\rm{total}}/ \frac{4}{3}\pi R_{0.5}^3$
for $\overline{R}\in\lbrace{0.05,0.10,0.20}\,\rm{pc}\rbrace$. \label{tab:density}}
\smallskip
\begin{tabular}{l l r r r} 
\hline
$R_{0.5,\rm{initial}}$ & $\overline{\rho}_{0.5,\rm{initial}}$ & $R_{0.5,\rm{final}}$ &
$\overline{\rho}_{0.5,\rm{final}}$ & $\overline{\rho}_{0.5,\rm{initial}}$ \\
$/\rm{pc}$ & $/M_{\odot}\rm{pc}^{-3}$ & $/\rm{pc}$ &
$/M_{\odot}\rm{pc}^{-3}$ & $/ \overline{\rho}_{0.5,\rm{final}}$ \\
\hline 
0.039 & $8.2 \times 10^5$ & 0.50 & 350 & 2400 \\ 
0.078 & $1.0 \times 10^5$ & 0.49 & 380 &  270 \\ 
 0.16 & $1.3 \times 10^4$ & 0.44 & 580 & 22 \\
\hline
\end{tabular}
\end{center}
\end{table*}

Our models have initial half-mass densities in the range $10^3 <
\overline\rho_{0.5}/M_{\odot}\,{\rm pc}^{-3} < 10^7$. The initial densities for
the smallest clusters seem a little extreme but there is increasing
evidence that the initial densities of open clusters are higher than
previously thought \citep{2009MNRAS.397.1577P}, that bound clusters can
expand quickly \citep{2008MNRAS.389..223B,2010MNRAS.404..721M}
and that rapid expansion can occur in the core
\citep{2001MNRAS.321..699K}.  Indeed Table~\ref{tab:density} shows that
the densest clusters expand the most because of the longer
dynamical time required and, interestingly, end up with very similar
half-mass radii to clusters that were initially somewhat sparser.
The final
densities vary because the initially denser clusters lose more mass
throughout the model. There is thus some uncertainty in
extrapolating back in time to estimate an initial cluster scaling.

\subsection{Collision frequency}
\label{seccollfreq}

\begin{table*}
\begin{center}
  \caption{Collision statistics for the models}
\smallskip
  \begin{tabular}{lrcccc}
    \hline
    & \multicolumn{2}{c} {Mean number of collisions}& & \multicolumn{2}{c}{Standard deviation} \\
    $\overline{R}/\text{pc}$& \multicolumn{1}{c}{preMS}& \multicolumn{1}{c}{ZAMS}&Ratio& \multicolumn{1}{c}{preMS}& \multicolumn{1}{c}{ZAMS} \\ \hline
    0.02&15.6&7.4&2.1&3.22&2.17\\
    0.05&8.1&2.9&2.8&1.73&1.73\\
    0.1&3.3&1.4&2.4&1.63&1.07\\
    0.2&1.8&0.9&2.0&1.62&0.88\\
    0.4&0.6&0.3&2.0&0.84&0.48\\ \hline
  \end{tabular}
  \label{tab:meancoll}
\end{center}
\end{table*}

\begin{figure}
  \begin{centering}
    \includegraphics[angle=-90,width=84mm]{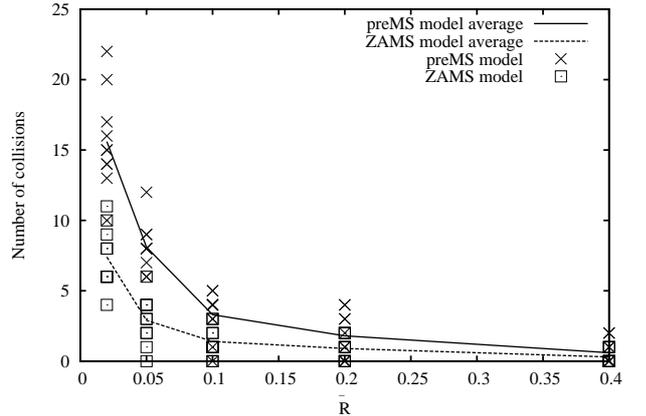}
  \end{centering}
  \caption{Number of collisions and their averages for 10 runs with and without
preMS evolution for
$\overline{R}=0.8R_{0.5}\in\{0.02,0.05,0.1,0.2,0.4\}\,\text{pc}$.}
  \label{fig:collisions}
\end{figure}

First we looked at the number of collisions between stars for each
$\overline{R}$ with and without preMS evolution. Averaged over ten
simulations, the results are listed in Table~\ref{tab:meancoll}, while
Figure~\ref{fig:collisions} shows the data and the averages over the
100 models.  As expected, more stars collided in the initially denser
clusters, as did more of the stars evolved from the top of their preMS
tracks.  The stars which collided in the ZAMS models had $R\approx
2R_{\odot}$ whereas in the preMS runs the radii were at least twice
this when they collided.

The gravitational focusing cross-section for
stars with masses $M_1$ and $M_2$, and radii $R_1$ and $R_2$ is
\begin{equation}
\sigma(M_1,M_2)=\pi(R_1+R_2)^2\left(1+\frac{2G(M_1+M_2)}{v_{\rm{rel}}^2(R_1+R_2)}\right),
\end{equation}
where $v_{\rm{rel}}$ is the relative velocity of the two stars at infinity
or at apocentre.  Because $v_{\rm{rel}}$ is usually small, this reduces to
\begin{equation*}
\sigma(M_1,M_2)\propto (R_1+R_2).
\end{equation*}
Thus doubling the radii of the two stars doubles the collision
cross-section and leads to roughly twice as many collisions.

To determine whether we are justified in ignoring the
pre-mainsequence evolution of stars over $8\,\rm M_\odot$ we check
the collision rate for such stars in our models.  For our
densest clusters with a half-mass radius of $0.02\,$pc the stellar
number density $n\approx 9\times 10^{-16}\,\rm R_\odot^{-3}$ and the typical
relative velocity $v_{\rm rel}\approx 11.5\,\rm km\,s^{-1}\approx
520\,R_\odot\,yr^{-1}$.  We estimate the collision cross-section by
setting $R_1 + R_2 = 100\,\rm R_\odot$ and $M_1 + M_2 = 10\,\rm
M_\odot$ and so find $\sigma \approx 9\times 10^6\,\rm R_\odot^2$.
From this we estimate a collision time-scale
\begin{equation}
t_{\rm coll} \approx \frac{1}{n\sigma v_{\rm rel}} \approx 2.4\times
10^5\,\rm yr.
\end{equation}
This is just a little shorter than the entire pre-mainsequence
lifetime of $3\times 10^5\,$yr for our $8\,\rm M_\odot$.  We recall that
it actually spends only $1.25\times 10^4\,$yr above $40\,\rm
R_\odot$.  So we can expect stars more massive than $8\,\rm M_\odot$
to contract to the main sequence before they collide again, but only
just in our densest cluster models.  If we were to model higher
densities we would need to model the pre-mainsequence evolution of
more massive stars.

\subsection{Runaways}

\begin{table*}
\begin{center}
    \caption{The ten most massive runaways \label{tab:runaways1}}
\smallskip
    \begin{tabular}{ccccrlr}
      \hline
      Model&$\overline{R}/\text{pc}$& $M_{\rm{initial}}/M_{\odot}$&
$M_{\rm{final}}/M_{\odot}$&$\Delta M/M_{\odot}$&Time/Myr$^{1}$&Number of collisions\\ \hline
      preMS &0.02 & 3.9 & 34.2\phantom{$^2$} & 30.3 & 3.37 & 18 \\
      preMS &0.02 & 3.6 & 30.6$^2$           & 27.0 & 2.56 & 12 \\
      preMS &0.02 & 2.1 & 29.9\phantom{$^2$} & 27.8 & 5.62 & 15 \\
      preMS &0.02 & 3.4 & 26.1\phantom{$^2$} & 22.7 & 1.21 & 13 \\
      preMS &0.02 & 3.5 & 22.1\phantom{$^2$} & 18.6 & 7.4 & 11 \\
      preMS &0.02 & 3.2 & 22.0\phantom{$^2$} & 18.8 & 2.01 & 11 \\
      preMS &0.05 & 3.2 & 21.6\phantom{$^2$} & 18.4 & 5.61 & 7 \\
      preMS &0.02 & 3.7 & 21.0\phantom{$^2$} & 17.3 & 9.57 & 9 \\
      preMS &0.05 & 3.7 & 20.9\phantom{$^2$} & 17.2 & 5.09 & 8 \\
      preMS &0.02 & 3.8 & 20.8\phantom{$^2$} & 17.0 & 5.62 & 14 \\ \hline
      ZAMS  &0.02 & 2.8 & 21.1\phantom{$^2$} & 18.3 & 1.59 & 8 \\
      ZAMS  &0.02 & 2.7 & 20.6\phantom{$^2$} & 17.9 & 1.99 & 11 \\
      ZAMS  &0.02 & 3.2 & 20.4\phantom{$^2$} & 17.2 & 8.22 & 7 \\
      ZAMS  &0.05 & 3.9 & 20.0\phantom{$^2$} & 16.1 & 2.77 & 5 \\
      ZAMS  &0.02 & 3.3 & 18.4\phantom{$^2$} & 15.1 & 2.19 & 5 \\
      ZAMS  &0.05 & 3.6 & 16.8\phantom{$^2$} & 13.2 & 0.85 & 4 \\
      ZAMS  &0.02 & 3.6 & 16.7\phantom{$^2$} & 13.1 & 3.71 & 4 \\
      ZAMS  &0.02 & 2.9 & 15.7\phantom{$^2$} & 12.8 & 3.85 & 5 \\
      ZAMS  &0.05 & 3.6 & 15.2\phantom{$^2$} & 11.6 & 4.69 & 4 \\
      ZAMS  &0.02 & 3.6 & 13.5\phantom{$^2$} & 9.9 & 6.12 & 5 \\ \hline
    \end{tabular}
\medskip
\par
$^1$Time is the age of the cluster when the runaway star collided for the last time.
\par
$^2$This star later evolved to a $6.8\,M_{\odot}$ black hole by $8.5\,$Myr
and absorbed another preMS star.
\end{center}
\end{table*}

In nearly every case, these collisions involved one or two stars
colliding many times. We call a star that collides with another more
than once a \emph{runaway}. These were usually the most massive stars
in the initial configuration, with initial masses $3\le
M_{\rm{initial}}/M_{\odot} \le 4$. Some gained over ten times
their initial mass within $10\,\rm{Myr}$ through multiple collisions.
Table~\ref{tab:runaways1} is a list of the ten most massive runaways
in both preMS and ZAMS models.  This shows that not only do more
collisions occur in the preMS models but also that the runaways end
up somewhat more massive.  The mean mass gained by runaways in
the preMS models was $12.0\,M_{\odot}$ (averaged over 34 runaways) and
$8.8\,M_{\odot}$ (24 runaways) in the ZAMS models.

\citet{2002ApJ...576..899P} found a similar
phenomenon in their rather extreme models of
colliding MS stars in dense star clusters where MS/MS collisions typically
form \emph{one} runaway star.  In Monte Carlo stellar dynamical models
\citet{2006MNRAS.368..141F}, while investigating a runaway mechanism
to create an intermediate-mass black hole in compact stellar clusters,
also found that only one runaway object was formed.  Our results are in line
with these findings.  Thus multiple collisions between preMS
stars early on in cluster evolution can be a viable mechanism to
create a few massive stars in otherwise low-mass clusters but is
unlikely to be the means to populate the top of the IMF
\citep{2007ARA&A..45..481Z}.  This conclusion was reached by
\citet{baumgardt2011} who also found typically only one runaway massive
star in their simulated clusters.

We found that the characteristic time to the first runaway is
about the same as that to form a small core, $0.05\,\rm{pc}$ in radius.  The
collisions, or merges, then occur within this core.  They also usually take
place at the pericentre of highly eccentric binary orbits.  In a few
cases a fly-by was seen to induce collisions between two stars in an
eccentric binary system.

Usually the most massive runaway stars formed in
the very dense $\overline{R}=0.02\,\rm{pc}$ model.  For $\overline{R}=0.4$ no
runaway stars formed at all and only three formed in the preMS model when
$\overline{R}=0.2$.  However there is significant variation between
the models.  In
the two least dense models the effect of the preMS evolution is
insignificant because the clusters are sparse enough
that by the time two stars come sufficiently close to collide the preMS
phase is over.

\section{Conclusions}
\label{conc}

We have not attempted to create a completely realistic
model of a cluster. The background gas that would exist in a
protostellar cloud has been neglected, as have the effects of
accretion of this gas \citep[see the treatment
by][]{baumgardt2011}.  Nor have we modelled the effects
of stellar accretion discs which would enhance the likelihood of
collisions and also change their nature.  Our aim here has been to
show that preMS evolution can have important consequences and to
identify when the preMS phase should be considered in more detail
in the future.

\begin{figure}
  \begin{centering}
\includegraphics[width=84mm]{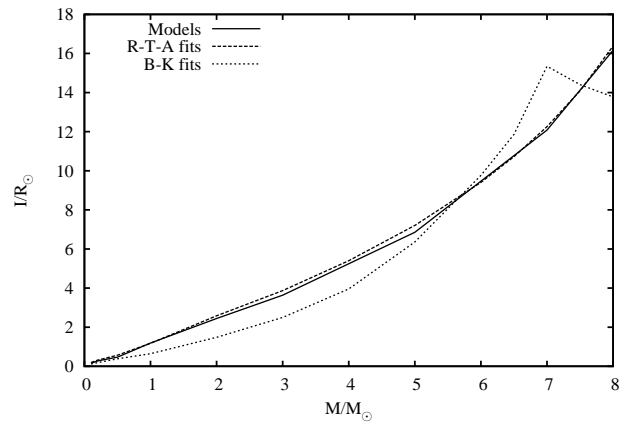}
\end{centering}
\caption{The integrated cross-section $I=\int_{\tau_{\rm preMS}}
R\,\mathrm{d}\tau$, taken over the preMS lifetime given by
equation~(\ref{eq:tzams}).  The models here are those of
\citet{tout1999}.  The fits of \citet{baumgardt2011} are
integrated from the time their accretion phase stops to when their
models reached the ZAMS.  Our cross-sections are relatively larger for
the longer lived lower-mass stars but smaller between about 5.5~and
$7.5\,\rm M_\odot$.}
\label{fig:xsection}
\end{figure}

It is not straightforward to make a direct comparison with
\citet{baumgardt2011} because they let their stars form by accretion
on to $0.1\,M_{\odot}$ cores, at constant accretion rates to populate
the IMF, whereas we start with an IMF and evolve our protostars as if
they were coeval.  However, examining the time-integrated
cross-section of our analytic fits compared to theirs (see
Figure~\ref{fig:xsection}) shows that their models systematically
underestimate the radii of the preMS stars below $6\,M_\odot$ compared
to ours.  This is important when looking at collision frequencies.
Calculating the time and mass-integrated cross-section
\begin{equation}
I_{\rm{IMF}}=\int _{0.1\,M_\odot}^{4.0\,M_\odot} \int_{\tau_{\rm preMS}} \eta(M) R(M,\tau)\,\mathrm{d}\tau\,\mathrm{d}M,
\end{equation}
where $\eta(M)$ is the Kroupa IMF normalised so that
\begin{equation}
\int _{0.1\,M_\odot}^{4.0\,M_\odot}\eta(M)\,\mathrm{d}M = 1,
\end{equation}
we find that for the original models $I_{\rm{IMF}}=11.5$, for our
analytic fits $I_{\rm{IMF}}=12.9$ and for the fits of
\citet{baumgardt2011}, $I_{\rm{IMF}}=7.9$.  Though we should therefore
expect to find more collisions in the preMS regime than they do the
differences in the models make it almost impossible to identify
whether this is actually the case.

The inclusion of gas is a natural next step in these models.  We would
expect an interplay between gas expulsion, which decreases the
star formation efficiency, accretion rates and the effect of gas
accretion on to stars.  Gas accretion not only leads to larger stars
and hence larger collision cross-sections but, in clusters, forces
contraction of the whole system and thence leads to more merging stars
\citep{1998MNRAS.298...93B,2002MNRAS.336..659B}, an increase in the
number of binary stars and early mass segregation
\citep{2011MNRAS.410.2799M}.

We would also expect our preMS stars to form discs while accreting or
indeed as a result of tidal disruption during a collision
\citep{2006MNRAS.370.2038D}.  The effect of such protoplanetary discs
on the number of collisions, their outcome and the time-scale on which
they take place should also be modelled in some way and
included in a more detailed $N$-body model.  Star--disc interactions
can lead to the disc being stripped away
\citep{disc,2001MNRAS.325..449S} or the formation of binary stars
\citep{1993MNRAS.261..190C} and, because many massive stars live in
tight binary systems \citep{2002ASPC..267..209Z}, it will be important
to know whether the preMS phase contributes to this.

In conclusion, although there is much physics that we have not
included, our models show that preMS evolution
increases both the number of collisions, when the density is
sufficiently large, and the amount of mass gained by the final
collision remnants.  Most collisions are part of a runaway
process so preMS collisions alone are probably insufficient to
populate the upper IMF. We identify an initial half-mass cluster
density of $10^4\,M_{\odot}\rm{pc}^{-3}$ below which preMS evolution
can be neglected because collisions are sufficiently rare.  However
caution should be exercised owing to large uncertainty
in the early length scales of clusters.

\section*{Acknowledgements}
CAT is very grateful to Churchill College for his Fellowship.  We
thank the referee for his helpful suggestions for improvement.


\begin{thebibliography}{36}
\expandafter\ifx\csname natexlab\endcsname\relax\def\natexlab#1{#1}\fi

\bibitem[{{Aarseth}(1999)}]{aarseth1999}
Aarseth, S.~J. 1999, PASP, 111, 1333

\bibitem[{{Bastian} {et~al.}(2008)}]{2008MNRAS.389..223B}
{Bastian}, N., {Gieles}, M., {Goodwin}, S.~P., {Trancho}, G.,
{Smith}, L.~J., {Konstantopoulos}, I., \& {Efremov} Y.
2008, MNRAS, 389, 223

\bibitem[{{Baumgardt} \& {Klessen}(2011)}]{baumgardt2011}
{Baumgardt}, H. \& {Klessen}, R.~S. 2011, MNRAS, 413, 1810

\bibitem[\protect\citeauthoryear{Bernasconi \&
    Maeder}{1996}]{bernasconi1996} Bernasconi, P. A. \& Maeder, A. 1996,
  A\&A, 307, 829

\bibitem[{{Benz} \& {Hills}(1987)}]{1987ApJ...323..614B}
{Benz}, W. \& {Hills}, J.~G. 1987, ApJ, 323, 614

\bibitem[{{Bonnell} \& {Bate}(2002)}]{2002MNRAS.336..659B}
{Bonnell}, I.~A. \& {Bate}, M.~R. 2002, MNRAS, 336, 659

\bibitem[{{Bonnell} {et~al.}(1997)}]{1997MNRAS.285..201B}
{Bonnell}, I.~A., {Bate}, M.~R., {Clarke}, C.~J., \& {Pringle}, J.~E.
1997, MNRAS, 285, 201

\bibitem[{{Bonnell} {et~al.}(1998){Bonnell}, {Bate}, \&
  {Zinnecker}}]{1998MNRAS.298...93B}
{Bonnell}, I.~A., {Bate}, M.~R., \& {Zinnecker}, H. 1998, MNRAS, 298, 93

\bibitem[{{Clarke} \& {Pringle}(1993)}]{1993MNRAS.261..190C}
{Clarke}, C.~J. \& {Pringle}, J.~E. 1993, MNRAS, 261, 190

\bibitem[{{Davies} {et~al.}(2006)}]{2006MNRAS.370.2038D}
{Davies}, M.~B., {Bate}, M.~R., {Bonnell}, I.~A., {Bailey}, V.~C.,
\& {Tout}, C.~A. 2006, MNRAS, 370, 2038

\bibitem[{{Davies} {et~al.}(1993)}]{1993A&A...272..430D}
{Davies}, M.~B., {Ruffert}, M., {Benz}, W., \& {Muller} E.
1993, A\&A, 272, 430

\bibitem[{{Eggleton}(1971)}]{eggleton}
{Eggleton}, P.~P. 1971, MNRAS, 151, 351

\bibitem[{{Freitag} {et~al.}(2006){Freitag}, {G{\"u}rkan}, \&
  {Rasio}}]{2006MNRAS.368..141F}
{Freitag}, M., {G{\"u}rkan}, M.~A., \& {Rasio}, F.~A. 2006, MNRAS, 368, 141

\bibitem[{{Hall} {et~al.}(1996){Hall}, {Clarke}, \& {Pringle}}]{disc}
{Hall}, S.~M., {Clarke}, C.~J., \& {Pringle}, J.~E. 1996, MNRAS, 278, 303

\bibitem[{{Hayashi}(1961)}]{1961PASJ...13..450H}
{Hayashi}, C., 1961 PASJ, 13, 450

\bibitem[{{Hayashi} \& {Nakano}(1963)}]{1963PThPh..30..460H}
{Hayashi}, C. \& {Nakano}, T. 1963, Progress of Theoretical Physics, 30, 460

\bibitem[{{Henyey} {et~al.}(1955){Henyey}, {Lelevier}, \&
  {Lev{\'e}e}}]{1955PASP...67..154H}
{Henyey}, L.~G., {Lelevier}, R., \& {Lev{\'e}e}, R.~D. 1955, PASP, 67, 154

\bibitem[{{Hillenbrand} \& {Hartmann}(1998)}]{1998ApJ...492..540H}
{Hillenbrand}, L.~A. \& {Hartmann}, L.~W. 1998, ApJ, 492, 540

\bibitem[{Hurley, Pols, \& Tout}(2000)]{hurley2000}
Hurley, J. R., Pols, O. R., \& Tout, C. A. 2000, MNRAS, 315, 543

\bibitem[{Hurley, Tout, \& Pols}(2002)]{hurley2002}
Hurley, J. R., Tout, C. A., \& Pols, O. R. 2002, MNRAS, 329, 897

\bibitem[{{Kroupa}(2001)}]{Kroupa01}
{Kroupa}, P. 2001, MNRAS, 322, 231

\bibitem[{{Kroupa} {et~al.}(2001){Kroupa}, {Aarseth}, \&
  {Hurley}}]{2001MNRAS.321..699K}
{Kroupa}, P., {Aarseth}, S., \& {Hurley}, J. 2001, MNRAS, 321, 699

\bibitem[{{Kumar}(1963)}]{1963ApJ...137.1121K}
{Kumar}, S.~S. 1963, ApJ, 137, 1121

\bibitem[{{Laycock} \& {Sills}(2005)}]{2005ApJ...627..277L}
{Laycock}, D. \& {Sills}, A. 2005, ApJ, 627, 277

\bibitem[{{McKee} \& {Tan}(2003)}]{2003ApJ...585..850M}
{McKee}, C.~F. \& {Tan}, J.~C., 2003, ApJ, 585, 850

\bibitem[{{Moeckel} \& {Bate}(2010)}]{2010MNRAS.404..721M}
{Moeckel}, N. \& {Bate}, M.~R. 2010, MNRAS, 404, 721

\bibitem[{{Moeckel} \& {Clarke}(2011)}]{2011MNRAS.410.2799M}
{Moeckel}, N. \& {Clarke}, C.~J. 2011, MNRAS, 410, 2799

\bibitem[{{Parker} {et~al.}(2009)}]{2009MNRAS.397.1577P}
{Parker}, R.~J., {Goodwin}, S.~P., {Kroupa}, P.,
\& {Kouwenhoven}, M.~B.~N. 2009, MNRAS, 397, 1577

\bibitem[{{Plummer}(1911)}]{1911MNRAS..71..460P}
{Plummer}, H.~C., 1911 MNRAS, 71, 460

\bibitem[{{Pols} {et~al.}(1995)}]{STARS}
{Pols}, O.~R., {Tout}, C.~A., {Eggleton}, P.~P., \& {Han}, Z.
1995, MNRAS, 274, 964

\bibitem[{{Portegies Zwart} \& {McMillan}(2002)}]{2002ApJ...576..899P}
{Portegies Zwart}, S.~F. \& {McMillan}, S.~L.~W. 2002, ApJ, 576, 899

\bibitem[{{Portegies Zwart} {et~al.}(2010)}]{2010ARA&A..48..431P}
{Portegies Zwart}, S.~F., {McMillan}, S.~L.~W.,
\& {Gieles}, M. 2010, ARA\&A, 48, 431

\bibitem[{{Scally} \& {Clarke}(2001)}]{2001MNRAS.325..449S}
{Scally}, A. \& {Clarke}, C. 2001, MNRAS, 325, 449

\bibitem[{{Stahler} \& {Palla}(2005)}]{2005fost}
{Stahler}, S.~W. \& {Palla}, F. 2005,
{The Formation of Stars} (Weinheim: Wiley-VCH)

\bibitem[{{Stahler} {et~al.}(2000){Stahler}, {Palla}, \&
  {Ho}}]{2000prpl.conf..327S}
{Stahler}, S.~W., {Palla}, F., \& {Ho}, P.~T.~P. 2000, in
Protostars and Planets IV, ed. V. Mannings, A. P. Boss, \& S. S. Russell
(Tucson: Univ. Arizona Press), 327

\bibitem[{{Tout} {et~al.}(1999){Tout}, {Livio}, \& {Bonnell}}]{tout1999}
{Tout}, C.~A., {Livio}, M., \& {Bonnell}, I.~A. 1999, MNRAS, 310, 360

\bibitem[{{Tout} {et~al.}(1996)}]{Tout96}
{Tout}, C.~A., {Pols}, O.~R., {Eggleton}, P.~P., \& {Han} Z.
1996, MNRAS, 281, 257

\bibitem[{{Zinnecker} \& {Bate}(2002)}]{2002ASPC..267..209Z}
{Zinnecker}, H. \& {Bate}, M.~R. 2002, in ASP Conf. Ser. 267, Hot Star
  Workshop III: The Earliest Phases of Massive Star Birth,
ed. P. Crowther (San Francisco: ASP), 209

\bibitem[{{Zinnecker} \& {Yorke}(2007)}]{2007ARA&A..45..481Z}
{Zinnecker}, H. \& {Yorke}, H.~W. 2007, ARA\&A, 45, 481

\end{thebibliography}
\end{document}